# Design of an achromatic, high numerical aperture optical assembly with a solid immersion lens

**JORDI PUIG,[1] CHRISTOPHE GALLAND,[1*]**

*1Laboratory of Quantum and Nano-Optics, Institute of Physics, Ecole Polytechnique Fédérale de Lausanne (EPFL), CH-1015 Lausanne, Switzerland*
*\*Corresponding author: chris.galland@epfl.ch*

**Abstract:** Efficiently collecting light emitted or scattered from nanoscale systems, which can be embedded in a high-index medium, is a challenge for fundamental spectroscopic studies and commercial applications of quantum dots, color centers, single molecules, plasmonic antennas, etc. Solid immersion lenses offer a powerful solution to increase the numerical aperture of the optical system, while being robust and compatible with cryogenic temperatures. However, they suffer from strong chromatic aberrations, limiting their applications to narrow-band excitation and emission spectral windows. Here, we design and optimize an achromatic lens assembly based on a solid immersion lens and a cemented doublet for broadband confocal spectroscopy. We provide an example with an achromatic range between 600 and 750 nm and show that it achieves diffraction limited spot size in excitation and in collection (reimaging of a point source) over more than 100 nm bandwidth. We expect our results to be useful to the broad community involved in spectroscopy and technological applications of nano-emitters for single photon sources, quantum information processing and quantum sensing.

## 1. Introduction

Confocal spectroscopy is a powerful tool to study the optical properties, the electronic and vibrational dynamics of systems whose dimensions are smaller than the wavelength of light (typically sub-100 nm). The emission or scattering pattern from these natural (e.g. molecules) or artificial nanostructures typically follows that of a point dipole. Depending on the orientation of this dipole, and the presence of a sharp interface between the embedding index of refraction and that of air, it can be challenging to achieve good excitation and collection efficiencies [1]. For example, not more than a few percent of the light emitted by a nitrogen-vacancy center in diamond can be collected from an external objective. Immersion lenses are a solution to this problem, by increasing the numerical aperture (NA) defined as $\text{NA} = n \cdot \sin(\theta)$ where n is the refractive index of the immersion medium and theta the maximum angle of acceptance with respect to the optical axis. At room-temperature and in a research setting, oil immersion offers a widely used alternative, with n ~ 1.5. However, oil cannot be used at cryogenic temperatures and is not well suited to commercial application with integrated nanoscale light sources. For these reasons, solid immersion lenses (SIL) are valuable tools, and can provide even higher NA thanks to higher refractive index materials [2–4]. For example, SIL have been used to efficiently collect light from color centers in diamond [5–7] and silicon carbide [8], from quantum dots [9], molecules [10], carbon nanotubes [11], etc. SIL were also recently employed to improve super-resolution microscopy of single NV centers [15].

For experimentalists, a recurring challenge faced when using SILs is their very pronounced chromatic aberration, in particular the chromatic focal shift (change in focal plane as a function of wavelength). This can be particularly problematic when dealing with broad emission spectra (such as the NV center emission with its phonon sidebands), or when performing photoluminescence excitation spectroscopy, where the excitation laser is scanned over 100 nm or more to characterize the excited state resonances and the phonon sidebands associated with a localized emitter, for example. In [13], a system with NA=1.9 is obtained by using a SIL, and chromatic aberration is corrected using a non-periodic phase structure element. But a very narrow achromatic range is obtained, considering wavelength deviations of only 1 nm. Another challenge is to couple efficiently the light collected through the SIL into a single-mode fiber, which is often required for single photon counting or when the sample is placed inside a cryostat physically detached from the detection apparatus. In [14], a SIL is used within a 2-lens system for near-field optics studies. High NA (1.84) is obtained, although chromatic aberrations were not corrected and the performance of the whole system were not analysed for collection from a point-like emitter (but only for focusing of collimated monochromatic light).

Previous experimental works and design studies thus did not pay close attention to the complete lens assembly consisting of the SIL itself, the collimation lens and the refocusing lens (for coupling into a fiber, a spectrometer, a pinhole for spatial filtering, etc). In particular, how these different elements can be leveraged together to correct for chromatic aberrations and obtain the best imaging performance has not been studied so far. Here, we investigate, design and optimize a full confocal excitation-collection system including a SIL and a cemented doublet. While keeping a spherical surface for the SIL (for easy manufacturing) and using a simple cemented doublet for focusing and collection, we achieve a very wide achromatic range of nearly 100 nm in the visible – near-infrared region (two orders of magnitude larger than in [13]) while maintaining diffraction-limited performances up to an NA of 1.6 (better than what is achievable with expensive oil-immersion objectives). Our results thus demonstrate an inexpensive and versatile solution for efficient and broadband in- and out-coupling of light to and from a nano-emitter.

## 1.1 Solid immersion lenses (SIL)

There are two main kinds of SIL: Hemispherical and Weierstrass SIL. These two types are described by Born and Wolf [12], where it is shown that light can be focused without spherical aberration at only two points within a high-index sphere. These focal points are known as the *aplanatic points* of the sphere. The first one is located at its center. As can be seen in Figure 1a, the incoming rays arrive at normal incidence on the air-SIL interface, so that there is no refraction at the surface of the sphere. This is the basis for a hemispherical SIL, which is a half-spherical lens in contact with the sample. As it has no refractive power, it enhances the NA by a factor *n* only (the refractive index of the SIL material), and the limiting factor regarding the system's NA is the collection half-angle, imposed by the NA of the collecting lens after the SIL.

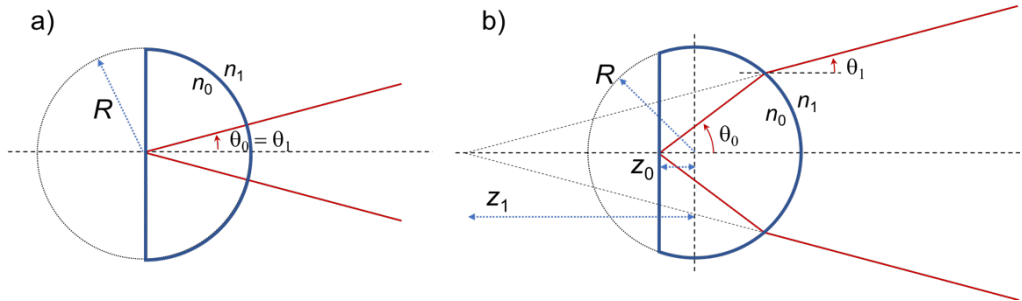

Fig. 1: a) Hemisphere and b) Weierstrass configurations for the use of solid immersion lens. Adapted from [4].

Weierstrass SIL exploit the second aplanatic point, which is located at a distance $z_0 = \left(\frac{n_1}{n_0}\right) \cdot R$ from the center of the sphere, where $R$ is the sphere radius and $n_0$ and $n_1$ are the refractive indices of the SIL and the air, respectively (Figure 1b). Light focused to this point, whose incident rays will undergo refraction at the air-SIL interface, has a virtual focus located outside the sphere at a distance $z_1 = \left(\frac{n_0}{n_1}\right) \cdot R$ from its center (Figure 1b). Due to this geometry, not only the emitter will be immersed in a high index material (NA enhancement by a factor $n_0$ compared to air), but light will also be refracted at the interphase (enhancement by another factor $n_0$ of the half-angle sinus due to Snell's law), both together resulting in an $n_0^2$ enhancement of the total NA. As a consequence, the generated diffraction-limited focal spot size will be smaller, and the collection efficiency will be improved.

### 1.2 Goal of this study

Our goal is to investigate the feasibility of an achromatic optical system based on a SIL and to propose an optimized design achieving an NA comparable to or higher than oil immersion objectives (NA>1.5). We will also study the field of the view of this optimized design to assess its use in imaging applications, where SIL have stringent limitations. The motivation for our work is the dearth of existing literature reporting achromatic optical systems including a SIL, be it theoretical or experimental. As stated in the introduction, there is an urgent need for both achromatic and high NA lens assemblies in spectroscopy and nano-photonics. For concreteness, we present the results for a SIL assembly whose achromatic wavelengths are 600 and 750 nm and which provides diffraction-limited performances with a collection NA of up to 1.6. The original Zemax files will be made available by the authors upon reasonable request to allow the reader to adapt the design to a particular application.

## 2. Methods

We now describe the methodology followed to achieve this goal.

### 2.1 Initial design for achromatic behavior

Optimization is a guided process that cannot explore an infinitely large configuration space. Before taking full advantage of the optimization tools described below, a starting point must therefore be provided whose behavior is sufficiently close to the targeted one. We used the lens-maker equations and solved them with the numerical software Matlab in order to design a lens assembly that would be achromatic in the desired range under the thin lens and paraxial approximations.

At least two optical elements with different dispersions should be combined in order to obtain achromatic behavior (two different wavelengths having the same focal distance). This is usually achieved with a strong positive lens made of low dispersion glass like crown glass, followed by a weaker negative lens made of high dispersion material like flint glass, glued together into a doublet. In this case the Lens Maker's formula is

$$\frac{1}{f_T(\lambda)} = \sum_i \frac{1}{f_i(\lambda)} = (n_1(\lambda)-1)\left(\frac{1}{R_1}-\frac{1}{R_2}\right) + (n_2(\lambda)-1)\left(\frac{1}{R_3}-\frac{1}{R_4}\right) \tag{1}$$

where $f_T(\lambda)$ is the doublet total focal length, $f_i(\lambda)$ is the focal length of each lens, $n_i(\lambda)$ the refractive index of each glass and $R_i$ successive surface curvature radiuses (where $R_1$ and $R_2$ correspond to the first and second surfaces on the first lens, and $R_3$ and $R_4$ correspond analogously to the second lens). The geometrical information can be simplified by the following notation:

$$\phi_1 = \left(\frac{1}{R_1}-\frac{1}{R_2}\right); \phi_2 = \left(\frac{1}{R_3}-\frac{1}{R_4}\right) \tag{2}$$

By choosing a desired $f_D$ and equaling the expression for the two desired wavelengths one obtains the following system of equations:

$$(n_1(\lambda_1)-n_1(\lambda_2))\cdot\phi_1 + (n_2(\lambda_1)-n_2(\lambda_2))\cdot\phi_2 = 0 \tag{3}$$

$$\frac{1}{f_D} = (n_1(\lambda_1)-1)\cdot\phi_1 + (n_2(\lambda_1)-1)\cdot\phi_2 \tag{4}$$

Once the system is solved, there is a free choice of the surface curvature radii to meet the obtained geometrical parameters, and the resulting assembly is achromatic for the wavelengths $\lambda_1$ and $\lambda_2$.

## 2.2 Ray tracing optimization

Using a starting point designed with the Lens Maker's equation, the subsequent lens designing was performed using the ray-tracing simulation software Zemax (version OpticStudio 17.5), which provides diverse tools for optimization and characterization of lens assemblies. In particular, the following tools have been used:

1. *Cross section*: 2-dimensional view of the lens system and proportions.

2. *Spot size diagram*: screen placed at the user's choice where the focus can be observed, the root mean square (RMS) spot radius is calculated and compared to the diffraction limited spot radius.

3. *Focal shift*: curve showing the shift in focal distance for different wavelengths, used to characterize chromaticity.

4. *Fields and wavelengths*: tool allowing to probe different wavelengths and fields. Fields stand for light sources that can be on or off axis, with different inclinations or angles. They allow to check the confocal performance in both excitation (where there is an incident collimated light beam) and collection (where the source is a point-like particle emitting in all directions).

5. *Aperture*: parameter that allows to choose the field's aperture. For incoming beams, it allows to choose the beam diameter, and for point sources, the numerical aperture or emission cone angle.

6. Optimization in Zemax is realized by minimizing a user-created merit function involving different weights. The merit function employed in the first stage of the optimization was a combination of the Axial Color (focal distance between two chosen colors), the RMS spot size and the System Focal Length, computed for different fields (e.g. beam incidence angle) and wavelengths. Finally, we used the Glass Substitution Hammer Optimization, which recommends a glass to substitute the selected one in order to better match the Merit Function requirements.

## 3. Results

### 3.1 Minimal assembly for achromaticity

Following the approach described above, we first experimented with the following configurations, which gave unsatisfactory results:

- *Triplet consisting of a Hemispherical SIL and an external doublet*. This configuration can be made achromatic using the external doublet, since the hemispherical SIL is not introducing any chromatic aberration. But since it has no refractive power, and the external doublet cannot be made with high numerical aperture, the maximum NA of the assembly that we achieved was 0.4, a value easily reached by standard microscope objectives without immersion.

- *Doublet consisting of a Weierstrass SIL and an external lens*. We could obtain a high collection cone angle (NA = 0.9) but not the achromatic behavior. This is due to the lack of degrees of freedom, as the Weierstrass SIL is a very constrained lens that focuses light in a specific point within its own spherical shape, so that changing the focal length of the first lens constrains the Weierstrass geometry, not allowing for the two hypothetical degrees of freedom that a doublet offers. We conclude that an achromatic optical system based a Weierstrass SIL requires, at least, another lens doublet in addition to the SIL.

Consequently, a minimum of 3 lenses, one Weierstrass SIL and a lens doublet, is required to fulfill the achromatic condition while obtaining a large NA>1, because the SIL does not add enough degrees of freedom to be part of just one achromatic doublet. The optical system obtained after optimization is shown in Fig. 2.

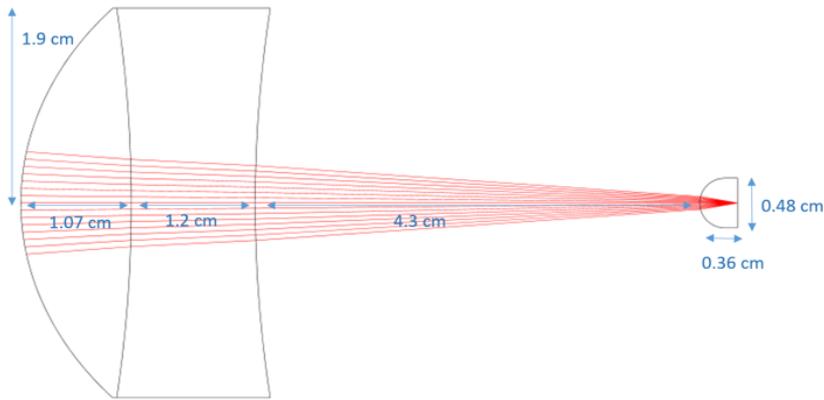

Figure 2: Layout and dimensions of the optimized design. Precise dimensions and all parameters of the optical surfaces are available as a Zemax file in the supporting material.

## 3.2 Optimized design for on-axis excitation and collection

The complete system, consisting of the Weierstrass SIL and cemented doublet (Fig. 2), was optimized to achieve the following performances:

1. Achromatic points (zero focal shift) at 600 and 750 nm (excitation geometry)
2. Diffraction limited spot in excitation
3. Collimated beam in collection (infinity-corrected imaging system)
4. Highest possible NA in detection

To optimize the system in excitation, a collimated laser beam (Gaussian profile) is launched from the left (incoming rays), with a variable diameter that determines the excitation NA (Fig. 3a). The RMS spot size is computed in the image plane of the Weierstrass SIL at different wavelengths. For a given NA and wavelength, Zemax also outputs the size of the Airy disc, which can be used as a benchmark for the focusing performance: we consider that any RMS spot radius below the Airy radius corresponds to a diffraction-limited spot size.

Evaluating the performance in detection requires more care, because assessing the quality of collimation is not straightforward. First, we model the emitter as a point source located on the image plane of the SIL, centered on the optical axis (see Section 3.4 for off-axis performances). Second, we place an ideal paraxial lens after the cemented doublet on the right (Fig. 3b), which reimages the point source on a screen. The ideal paraxial lens of Zemax focuses a perfectly collimated beam onto a single point. Therefore, analyzing the RMS spot size on the screen allows us to assess and optimize the performance of the optical system in detection. In practice, the reimaging system could be another lens assembly optimized separately. Our design is thus "infinity corrected", complying with modern standards in microscopy. We choose to optimize the system for the highest possible NA in detection, since this is what determines the overall detection efficiency.

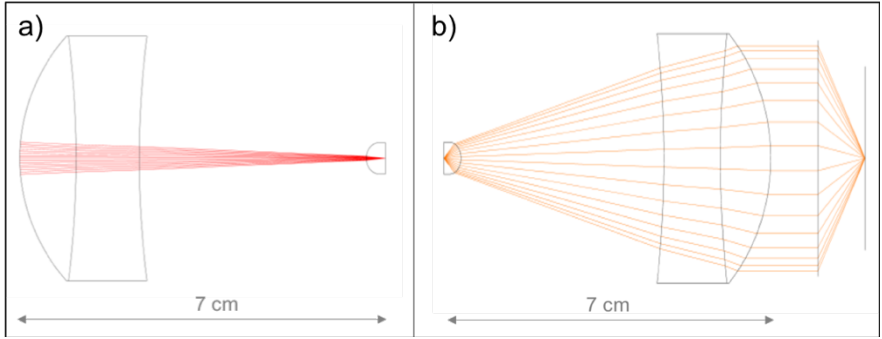

Figure 3: Excitation (a) vs. collection (b) configuration. Rays are being traced from left to right. In the collection configuration, the last lens is an ideal paraxial lens used to probe how well the beam is collimated. This lens can represent any imaging system placed before the detector of the optical fiber used in detection.

Finally, the achromatic range is defined by imposing that the chromatic focal shift is zero at both 600 nm and 750 nm (Fig. 4), when the system is excited with a collimated beam with 10 mm diameter.

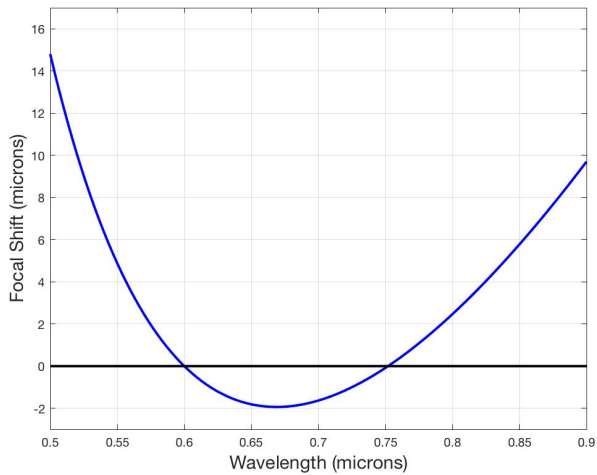

Figure 4. Focal shift along the optical axis for a wavelength range of 0.5-0.9 microns (blue line). The black line highlights the points with 0 focal shift. This study has been done using an excitation laser beam of 10 mm in diameter.

## 3.3 Performance of the optimized design for on-axis excitation and collection

We first present in Fig. 5 the detection performance of the system at a fixed wavelength of 700 nm while the NA is increased. Remarkably, a diffraction limited spot size is achieved up to an NA of 1.6, before strongly aberrated rays start degrading the imaging quality. Our system is thus on par with the state-of-the-art SIL-based confocal microscopes.

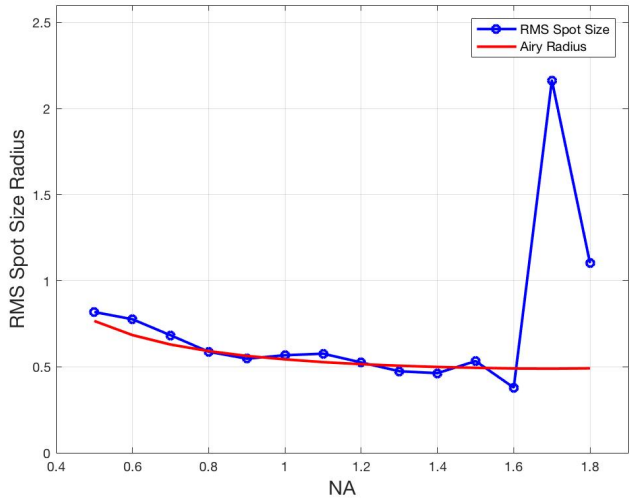

Figure 5. RMS spot radius (blue line) for different values of the detection NA, ranging from 0.5 to 1.8, at a fixed wavelength of 700 nm. The red line is the diffraction limit represented by the Airy disc radius. These values are obtained on the image plane of an ideal paraxial lens, which would focus a collimated beam into a single point.

Next, we show in Fig. 6 how the excitation and detection spot sizes depend on the wavelength, at a fixed NA (corresponding to a fixed beam diameter in excitation). These results confirm our expectation that the system performs close to the diffraction limit over a broad wavelength range of more than 100 nm.

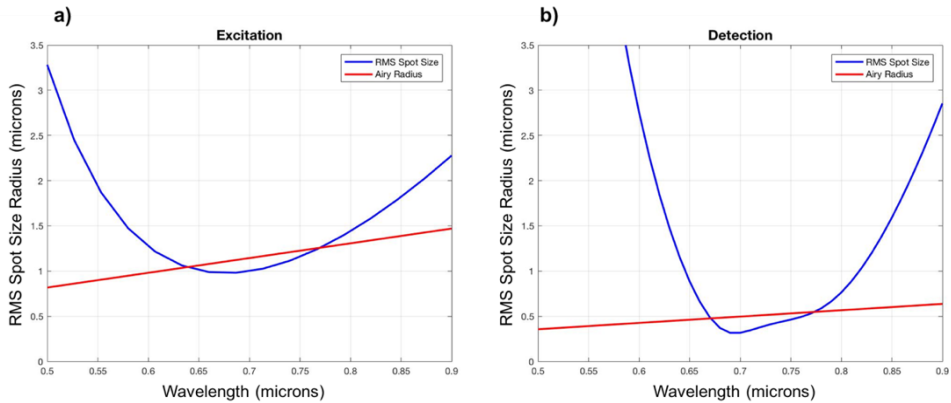

Figure 6. Plot a) shows the RMS spot radius obtained in excitation (blue line) for wavelengths ranging from 0.5 to 0.9 microns, and the diffraction limit or Airy radius (red line). For this plot, an excitation beam diameter of 10 mm (equivalent to NA = 0.56) was used. Plot b) shows the RMS spot radius obtained in detection on the reimaging plane (blue line) when focusing the collimated beam with an ideal non-aberrated paraxial lens. The Airy spot radius is shown with the red line. For this plot, a collection NA of 1.5 was used.

## 3.4 Characterization of field of view

After obtaining a lens assembly with the targeted numerical aperture, spot size and achromaticity, we performed a study of the field of view, to assess the suitability of the SIL for wide-field imaging applications. We computed the reimaging spot size in the collection geometry for different lateral displacements of the point source emitter relative to the optical axis, as illustrated in Fig. 7.

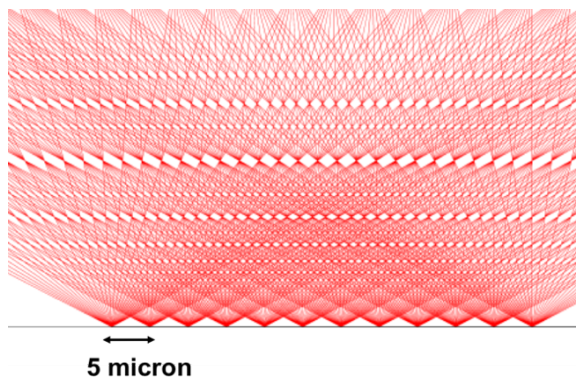

Figure 7 Schematics of field-of-view characterization

For each displacement, the RMS spot size on the ideal detector was computed, and the results are displayed in Fig. 8. Reducing the numerical aperture, as expected, helps increasing the field of view, but even for NA=1.0 it remains on the order of 10 μm radius around the optical axis. This is an intrinsic limitation of the SIL, which may be somewhat improved by using larger SIL radii.

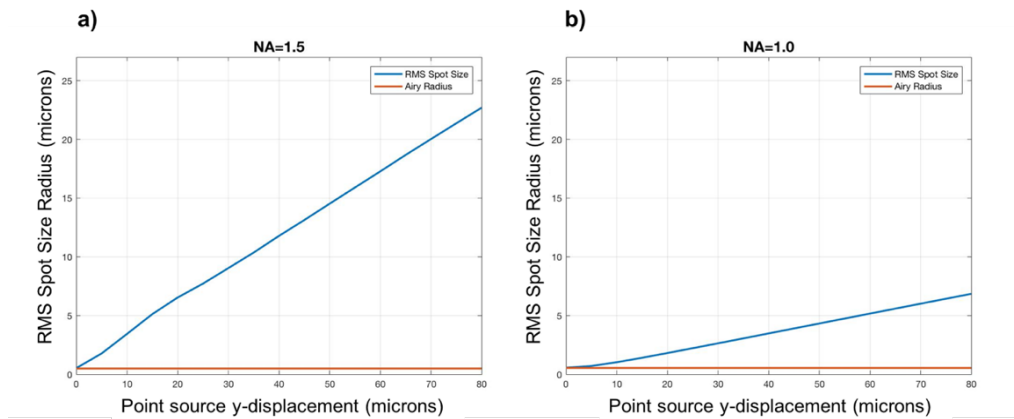

Figure 8: RMS spot size vs. lateral displacement of the point source for two different NA.

## 4. Conclusions

We designed and optimized a lens assembly for confocal spectroscopy with high numerical aperture (up to NA 1.6) and large achromatic range (600 – 750 nm) based on a solid immersion and a lens doublet, which achieves diffraction-limited performances along the optical axis. For off-axis emitters, the performance quickly degrades, as expected for a small solid immersion lens, limiting the field of view (for a fixed SIL) to a radius of about 10 μm when the NA is reduced to 1. We expect our design to be valuable for a wide range of applications in nano-emitter spectroscopy, quantum optics, nano-photonics and nano-plasmonics, where diffraction limited performance and achromatic behavior is required over a broad wavelength range of 100 nm or more, in excitation and/or collection.

## Funding

Swiss National Science Foundation (SNSF), Grant PP00P2_170684

## Acknowledgments

We thank the Laboratory of Applied Photonics Devices of Prof. Christophe Moser for enabling this project.

## Disclosures

The authors declare that there are no conflicts of interest related to this article.